\documentclass[preprint]{revtex4-1}

\usepackage[english]{babel}
\usepackage[math]{blindtext}
\usepackage{lineno}
\modulolinenumbers[5]
\usepackage{amssymb}
\usepackage{amsmath}
\usepackage{amsfonts}
\usepackage{slashed}
\usepackage{enumerate}
\usepackage{graphicx}
\usepackage{relsize}
\usepackage{color}
\usepackage{extarrows}

\definecolor{gray}{rgb}{0.5, 0.5, 0.5}

\begin{document}

\title{Symmergent Gravity, Seesawic New Physics, and their Experimental Signatures}
\author{Durmu{\c s} Demir}
\affiliation{Department of Physics, {\.I}zmir Institute of Technology, TR35430, {\.I}zmir, Turkey}

\begin{abstract}
The standard model  of elementary particles (SM) suffers from various problems, such as power-law ultraviolet (UV) sensitivity, exclusion of general relativity (GR),  and absence of a dark matter candidate. The LHC experiments, according to which the TeV domain appears to be empty of new particles, started sidelining TeV-scale SUSY and other known cures of the UV sensitivity. In search for a remedy, in this work, it is revealed that affine curvature can emerge in a way restoring gauge symmetries  explicitly broken by the UV cutoff. This emergent curvature cures the UV sensitivity and incorporates GR as symmetry-restoring emergent gravity ({\it symmergent gravity}, in brief) if a new physics sector (NP) exists to generate the Planck scale and if SM+NP is fermi-bose balanced. This setup, carrying fingerprints of trans-Planckian SUSY, predicts that gravity is Einstein (no higher-curvature terms), cosmic/gamma rays can originate from heavy NP scalars, and the UV cutoff might take right value to suppress the cosmological constant (alleviating fine-tuning with SUSY).  The NP does not have to couple to the SM.  In fact, NP-SM coupling can take any value from  zero to  $\Lambda^2_{SM}/\Lambda^2_{NP}$ if the SM is not to jump from $\Lambda_{SM}\approx 500\, {\rm GeV}$ to the NP scale $\Lambda_{NP}$. The zero coupling, certifying an undetectable NP, agrees with all the collider and dark matter bounds at present. The {\it seesawic} bound $\Lambda^2_{SM}/\Lambda^2_{NP}$, directly verifiable at colliders, implies that: {\it (i)} dark matter must have a mass $\lesssim \Lambda_{SM}$,  {\it (ii)} Higgs-curvature coupling must be $\approx 1.3\%$,  {\it (iii)} the SM RGEs must remain nearly as in the SM, and {\it (iv)} right-handed neutrinos must have a mass $\lesssim 1000\, {\rm TeV}$.  These signatures serve as a concise testbed for symmergence.  
\end{abstract}

\maketitle

\tableofcontents

\section{Introduction}
The SM, a spontaneously broken renormalizable quantum field theory (QFT) of the strong and electroweak interactions, has shown good agreement with all the experiments performed so far \cite{sm-sofar,exotica}. Its parameters have all been fixed experimentally. This does, however, not mean that it is a  complete theory. Indeed, it is plagued by enigmatic problems like destabilizing UV sensitivities \cite{veltman}, exclusion of gravity \cite{grav}, and absence of a  dark matter candidate \cite{cdm}, which are impossible to address without new physics beyond the SM (BSM). Schematically,
\begin{flalign}
\label{NP}
\text{BSM} = \text{gravity}\, ({\text{GR}})\, +\, {\text{new QFT beyond the SM}\, \text{(NP)}}
\end{flalign}
in which the general relativistic (GR) structure of gravity is revealed by various experiments \cite{will}  and observations \cite{lam-cdm}. The NP is under way at colliders \cite{exotica} and dark matter searches \cite{cdm-current}.

Quantum correction to the Higgs boson mass, quadratically sensitive to UV boundary \cite{veltman}, exceeds the Higgs boson mass just above the electroweak scale. This means that the SM must stop working below the TeV scale. It does not stop, however. Indeed, the LHC has confirmed the SM \cite{exotica} up to multi-TeV energies. This contradiction between the SM loops and the LHC can be eliminated only if the NP in (\ref{NP}) improves the SM in a way without introducing any new interacting particles. The present work approaches this puzzling requirement via proof-by-contradiction, that is, it begins by assuming that the NP is absent (Sec. II and Sec. III) and, at a later stage, it ends up with NP  through the consistency of the induced gravitational constant (Sec. IV).

The GR must be incorporated into the SM \cite{will,lam-cdm}. This has been attempted with classical GR \cite{cgr} (despite \cite{cgr2}), quantized GR  \cite{qgr1} (despite \cite{qgr2,bicep}), and emergent GR \cite{sakharov} (see also \cite{emerge-verlinde,emerge-raamsdonk,emerge-lit,terazawa}). The present work incorporates curvature into the SM effective action in flat spacetime (Sec. II) by building on the nascent ideas proposed in \cite{demir1,demir2} and subsequent developments voiced in \cite{talk-ppc, talk-NP, talk-hepac, talk-qdis}. It incorporates  gravity in a way restoring the gauge symmetries broken by the UV cutoff \cite{hard-break} (Sec. III B), in a way elasticating the rigid flat spacetime (Sec. III C), and in a way involving a nontrivial NP sector to induce the gravitational constant (Sec. IV). This gauge symmetry-restoring emergent gravity, {\it symmergent gravity} \cite{talk-NP, talk-hepac, talk-qdis} in brief, sets up a  framework (Sec. IV) in which
\begin{enumerate}
\item curvature arises as a manifestation of the elastication of the flat spacetime,

\item GR emerges along with the restoration of gauge invariance,

\item gravitational constant  necessitates an NP sector,

\item SM + NP possesses exact 
fermi-bose balance and the induced gravitational constant is suggestive of a trans-Planckian SUSY breaking \cite{susy}, 
\item the UV boundary can be fixed, in principle, by  suppression of the cosmological constant (with no immediate solution for the cosmological constant problem \cite{ccp} though)
\end{enumerate}
so that there arise a number of  descriptive signatures with which symmergence can be probed via decisive experiments (Sec. V):
\begin{enumerate}
\item Higher-curvature terms are predicted to be absent. This excludes, for instance, $f(R)$ gravity \cite{fR} and agrees well with the current cosmological data  \cite{lam-cdm}. 

\item NP scalars with trans-GZK \cite{GZK} VEVs are predicted to give cause to
cosmic rays \cite{cosmic-ray} (digluon) as well as gamma rays \cite{gamma-ray} (diphoton) via certain Planck-suppressed higher-dimension operators.

\item Symmergence does not necessitate any SM-NP coupling for it to work. This property, not found in the known SM completions (SUSY, extra dimensions, compositeness and others \cite{beyond}) for which a sizable SM-NP coupling is essential, provides a rationale for stabilizing the electroweak scale against the SM-NP mixing. Indeed,  if the SM-NP coupling goes like $m_H^2/m_{NP}^2$ then the Higgs boson mass $m_H$ remains within the allowed limits. This seesawic (seesaw-wise) structure leads to various  testable features:
\begin{enumerate}
    \item It is predicted that heavier the NP larger the luminosity needed to discover it. This distinctive feature can be probed at present \cite{coupling-LHC} and future colliders \cite{coupling-FCC,coupling-ILC}.   
    
    \item It is also predicted that the right-handed neutrinos \cite{neutrino-portal} must weigh below a 1000 {\rm TeV} (see also \cite{vissani}). This bound can be tested at future colliders \cite{biz} if not at the near-future SHiP experiment \cite{SHiP}.
    
    \item It turns out that the SM couplings (gauge and non-gauge) must run as if NP is absent if the NP lies sufficiently above the electroweak scale. This feature, which rests on the fact that symmergence leaves behind only logarithmic sensitivity to the UV boundary \cite{cutoff-dimreg},   can be tested at present \cite{coupling-LHC} and future colliders \cite{coupling-FCC,coupling-ILC}.
    
    \item Symmergence accommodates both ebony (having only gravitational interactions with the SM) \cite{demir2,ebony} and dark (having seesawic couplings to the SM)   matters. They both agree with the current bounds. The latter, thermal dark matter, is predicted to weigh below the electroweak scale. This agrees with current limits and can be tested further in future searches \cite{cdm-current,dm-LHC}. 
    
    \item It is predicted that, in the SM, non-minimal Higgs-curvature coupling equals $1.3 \%$ at one-loop and remains so unless the NP lies near the electroweak scale. This coupling, too small to drive the Higgs inflation \cite{Higgs-inflation}, can serve as a testbed at collider experiments \cite{non-min-cinli} if not in the astrophysical or cosmological environments.   
\end{enumerate}
\end{enumerate}
The work is concluded in Sec. VI.

\section{UV Boundary, Effective SM and the UV Sensitivity Problems}
The NP needed to complete the SM, roughly sketched
in (\ref{NP}), can be elucidated only after a complete  picture of the SM in regard to its UV boundary and UV sensitivity.  

\subsection{UV Boundary} 
The Higgs mass-squared $m_{H}^2$, measured to be $\left(m_H^2\right)_{LHC}\approx (125\, {\rm GeV})^2$ at the LHC \cite{higgs-mass}, is overwhelmed by the quantum correction \cite{veltman}
\begin{flalign}
\delta m_H^2 = c_H \Lambda^2 
%+ c^H_\digamma m_{\digamma}^2 \log\frac{m_\digamma}{\Lambda}
\label{deltmh-0}
\end{flalign}
in which  $\Lambda$ is the UV boundary of the SM, and
\begin{flalign}
c_H = \frac{3}{16 \pi^2 \left|\langle H \rangle \right|^2} \left(4 m_t^2 - 2 M_W^2 - M_Z^2 - m_H^2\right) \approx 5.14\times 10^{-2}
\label{veltman-factor}
\end{flalign}
is the loop factor. The correction (\ref{deltmh-0}), a one-loop SM effect, grows quadratically with $\Lambda$ and exceeds $\left(m_H^2\right)_{LHC}$ already at $\Lambda=\Lambda_{W}$, where
\begin{flalign}
\Lambda_{W} \approx 550\, {\rm GeV}
\end{flalign}
lies just above the electroweak scale $|\langle H \rangle| \approx 246.22\ {\rm GeV}$. This low-lying $\Lambda_W$, which can be changed slightly by incorporating subleading $\log m_H/\Lambda$ corrections to (\ref{deltmh-0}), is a characteristic feature of the SM spectrum and the experimental result  $\left(m_H^2\right)_{LHC}$. It implies that the SM must stop working at $\Lambda_{W}$. It does not stop, however. Indeed, the LHC experiments show that the SM continues to hold good up to multi-TeV energies without any new  field. This contradictory UV overextension is the problem.  There is no clear solution. There is even no clear way to search for a solution. There is, however, a possibility that a mechanism, not necessarily unique, might be constructed via proof-by-contradiction \cite{demir2,talk-NP}, that is, by first
\begin{flalign}
&{\textsf{constructing the SM effective action below}}\, \Lambda_{{W}}\, {\textsf{assuming that}}\nonumber\\
&{\textsf{there is nothing but the SM all the way up to}}\, \Lambda_{{U}}\gg \Lambda_{{LHC}},
\label{cond1}
\end{flalign}
and then 
\begin{flalign}
& {\textsf{revealing the necessity and structure of the NP for incorporation of}}\nonumber\\ &{\textsf{gravity and neutralization of the destabilizing}}\, \Lambda_{{U}}\, {\textsf{sensitivities.}}
\label{cond2}
\end{flalign}
The first step sets up a UV boundary $\Lambda_U$ and reveals SM's UV sensitivity. The second, on the other hand, uncovers NP via 
induction of gravity and fixes $\Lambda_U$ in terms of the NP scale. 

\subsection{SM Effective Action}
In accordance with (\ref{cond1}), integration of the fast modes (fields with energies $\gtrsim \Lambda_{{W}}$) out of the SM spectrum gives an effective action for slow  modes $\digamma$ (fields with energies $\lesssim \Lambda_{{W}}$) \cite{eff-ac}
\begin{flalign}
\label{action-flat}
S_{eff}\left(\eta\right) &= {S_{tree}\left(\eta, \digamma\right)} +  {\delta S_{log}\Big(\eta, \digamma, \log\frac{\Lambda_{{W}}}{\Lambda_{{U}}}\Big)}\nonumber\\&+ {\delta
 S_{O}\left(\eta, \Delta^2\right)}+ {\delta S_{H}\left(\eta, \Delta^2\right)}+ {\delta S_{V}\left(\eta, \Delta^2\right)}
\end{flalign}
in which $\eta_{\mu\nu}$ is the flat metric, $H$ is the slow Higgs field, $V_{\mu}$ are the slow gauge fields (photon, gluon, $W$ and $Z$), and finally
\begin{flalign}
\label{gap}
\Delta^2 = \Lambda_{{U}}^2- \Lambda_{{W}}^2
\end{flalign}
is the UV-EW gap. The tree-level SM action $S_{tree}$ and the logarithmic corrections $\delta S_{log}$ both lie below $\Lambda_{{W}}$. But, the other three 
\begin{flalign}
\delta S_{O}&= -\int d^4x \sqrt{-\eta}\, \Big\{\left(2 c_O \Lambda_{{W}}^2 + \sum_{\digamma} c_\digamma m_\digamma^2 \right) \Delta^2   + {c_O} \Delta^4 \Big\} \label{deltSO}\\
\delta S_{H}&= -\int d^4x \sqrt{-\eta}\, {c_H} \Delta^2\,  H^{\dagger} H\label{deltSH}\\
\delta S_{V}&= \int\! d^4x \sqrt{-\eta}\, {c_V} \Delta^2\, {\mbox{Tr}}\!\Big[ V_{\mu} V^{\mu}\Big]\label{deltSV}
\end{flalign}
pull the SM off the electroweak scale depending on how large $\Delta^2$ is. They tend to destabilize the SM and it is this destabilization that necessitates  a neutralization mechanism. Their Wilson coefficients $c_{O,\cdots, V}$  involve only the ratio \cite{veltman}
\begin{flalign}
\label{hierarchy}
 \frac{\Lambda_{{W}}}{\Lambda_{{U}}}
\end{flalign}
as the measure of the EW-UV hierarchy. 
%It is these UV-born quantum corrections that destabilize the SM in various channels. 

\subsection{UV Sensitivity Problems}
The power-law quantum corrections in (\ref{deltSO}), (\ref{deltSH}) and (\ref{deltSV}) give cause to serious destabilization problems. They are tabulated in Table \ref{table-prob}. The cosmological constant problem (CCP) \cite{ccp}, caused by $\delta S_{O}$, exists only when gravity is present.  The big hierarchy problem (BHP) \cite{veltman} (gauge hierarchy problem) refers to quadratic UV sensitivities of the Higgs (from  $\delta S_{H}$) and $W/Z$ (from $\delta S_{V}$) masses. The electric charge or color breaking (CCB) \cite{hard-break}, on the other hand, arises from the photon and the gluon mass terms in $\delta S_{V}$ (purely quadratic in $\Delta$). The SM is impossible to make sense in the UV before these problems are satisfactorily resolved.
\begin{table}[ht]
\caption{Quantum corrections and problems they give cause to. The coefficients $c_\digamma, \dots, c_g$ are loop factors that depend on  ${\Lambda_{{W}}}/{\Lambda_{{U}}}$. (CCP, with ${\tilde{c}}_O = c_O + (2 c_O \Lambda_W^2 + \sum_{\digamma} c_\digamma m^2_\digamma)/\Delta^2$, makes sense only in curved geometry hence the grey color.) \label{table-prob}}{% 
\begin{tabular}{r||l|l|l|l|l} % centered columns (4 columns)
\hline %inserts double horizontal lines
  & ${\delta S_{log}}$ & ${\delta S_{H}}$ & ${\delta S_{V}}$  & ${\delta S_{O}}$& Problem \\\hline\hline
$\delta V$ & $\neq 0$ & 0 & 0 &${\tilde{c}}_O \Delta^4$& {\color{gray}CCP}\\\hline
$\delta m_H^2$ & $\neq 0$ &${c_H} \Delta^2$&0&0& BHP\\\hline
$\delta m_W^2$ & $\neq 0$ &0&${c_W} \Delta^2$&0& BHP\\\hline
$\delta m_Z^2$ & $\neq 0$ &0&${c_Z} \Delta^2$&0& BHP\\\hline
$\delta m_\gamma^2$ &0&0&${c_\gamma} \Delta^2$& 0& CCB\\\hline
$\delta m_g^2$ &0&0&${c_g} \Delta^2$& 0& CCB\\\hline
\end{tabular}}
\end{table}

\subsection{Impossibility of Renormalizing Away CCP, CCB and BHP}
The SM is a renormalizable QFT. If so, why isn't it possible to include $\delta S_{log}$, $\delta S_{O}$, $\delta S_{H}$, $\delta S_{V}$ into a renormalization of $S_{tree}$ to get rid of the problems in  Table \ref{table-prob}? Because $\Delta$ is physical. Indeed,  $\Lambda_U$ can pertain to gravity or NP sectors \cite{veltman}. Moreover, there is simply no place to hide $\delta m_{\gamma,g}^2 = c_{\gamma,g} \Delta^2\neq 0$ since $(m_{\gamma,g}^2)_{tree} = 0$.  These problems are therefore physical and their solutions entail physical changes on the SM (like inclusion of gravity). 

\section{Incorporation of Curvature into the Effective SM}
It is now time to ascertain how curvature can be incorporated into the flat spacetime effective SM in (\ref{action-flat}) and how that incorporation affects the UV sensitivity problems in Table \ref{table-prob}. 

\subsection{How not to Incorporate: Curvature by Hand}
\label{subsec:notrestore}
Gravity is incorporated into classical field theories in flat spacetime by first mapping the flat metric $\eta_{\mu\nu}$ into a putative curved metric $g_{\mu\nu}$ as
\begin{flalign}
\label{map-metric}
\eta_{\mu\nu}\, {{\hookrightarrow}}\, g_{\mu\nu}
\end{flalign}
in view of general covariance \cite{eqp}, and then adding curvature of the Levi-Civita connection
\begin{flalign}
{}^{g}\Gamma^{\lambda}_{\mu\nu} = \frac{1}{2} g^{\lambda\rho}\left(\partial_{\mu} g_{\nu\rho} + \partial_{\nu} g_{\rho\mu} - \partial_{\rho} g_{\mu\nu}\right)
\end{flalign}
to make $g_{\mu\nu}$ dynamical. The curvature sector, added by hand, ignoring ghosts, can involve all curvature invariants \cite{h-curv}. It can involve, for instance, the Ricci tensor $R_{\mu\nu}\left({}^{g}\Gamma\right)$
in traced ($R = g^{\mu\nu} R_{\mu \nu}\left({}^{g}\Gamma\right)$), squared 
($R_{\mu\nu}\left({}^{g}\Gamma\right) R^{\mu\nu}\left({}^{g}\Gamma\right)$) or in any other invariant form.  This curvature-by-hand method, a standard procedure for classical field theories, leads to
\begin{flalign}
S_{eff}\Big(g\Big)  - \int d^{4}x  \sqrt{-g}
\left\{ {\tilde M}^2 R + {\tilde{\alpha}} R^2 + {\tilde{\beta}} R^{\mu\nu}  R_{\mu\nu}  + \cdots \right\}  \label{tilded-1} 
\end{flalign}
when applied to the SM effective action in (\ref{action-flat}).  The problem with this action is that ${\tilde{M}}$, ${\tilde{\alpha}}$, ${\tilde{\beta}}$, $\cdots$  are all inherently incalculable \cite{demir2,talk-NP,talk-hepac}. This is because matter loops have already been used up in forming the flat spacetime effective action $S_{eff}$, and there have remained thus no loops to induce any extra interaction, with or without curvature. This incalculability constraint, which reveals the difference between classical and effective field theories, renders the tentative action (\ref{tilded-1}) unphysical. It is in this sense that the general covariance \cite{eqp} is not adequate for incorporating curvature into effective SM. In essence, what is needed is a separate covariance relation between the {\it scales in $S_{eff}$} (say, $\Lambda_{{W}}^2$ or $\Delta^2$) and {\it curvature} \cite{talk-ppc}  so that gravity can be incorporated in a way involving no arbitrary, incalculable constants.

\subsection{How to Incorporate: Curvature by Gauge Symmetry Restoration}
\label{subsec:restore}
The gauge part $\delta S_{V}\left(\eta, \Delta^2\right)$, which  must be neutralized  for color and electromagnetism to remain exact and electroweak breaking to be spontaneous, poses a vexed problem due to strict masslessness of the photon and the gluon. It can be tackled via neither the Stueckelberg method \cite{stueckelberg} nor the spontaneous symmetry breaking \cite{ssb}. It can, nonetheless, be tackled by furthering the nascent ideas proposed in \cite{demir2} (which attempts at restoring gauge symmetry within the GR with fixed $\Lambda_{{U}}^2 + \Lambda_{{W}}^2$) and subsequent advancements voiced in \cite{talk-ppc, talk-NP, talk-hepac, talk-qdis}. 

\subsubsection{$\delta S_{V}\left(\eta, \Delta^2\right)$ in a New Light}
\label{subsubsec:restore1}
It proves useful to start with the obvious identity \cite{demir2,talk-ppc}
\begin{flalign}
\delta S_{V}\left(\eta, \Delta^2\right)= \delta S_{V}\left(\eta, \Delta^2\right) - I(\eta,V) + I(\eta,V) \label{1st}
\end{flalign}
in which the gauge-invariant kinetic construct
\begin{flalign}
I(\eta,V) = \int d^{4}x \sqrt{-\eta}
\frac{c_V}{2} {\mbox{Tr}}\left\{ \eta_{\mu\alpha} \eta_{\nu\beta}V^{\mu\nu} V^{\alpha\beta}\right\}
\end{flalign}
is subtracted from and added back to $\delta S_V$. This construct, involving the loop factor $c_V$ and the field strength tensor $V_{\mu\nu}$, leads to 
\begin{flalign}
\delta S_{V}\left(\eta, \Delta^2\right) &=
 - I(\eta,V)\nonumber\\ &+ \int d^{4}x  \sqrt{-\eta} {c_V} {\mbox{Tr}}\!\left\{V^{\mu}\left( -D_{\mu\nu}^2 + \Delta^2 \eta_{\mu\nu} \right)\! V^{\nu}\right\}\nonumber\\
 &+\int d^{4}x  \sqrt{-\eta} {c_V} {\mbox{Tr}}\!\left\{{\partial}_{\mu} \left(\eta_{\alpha\beta} V^{\alpha} V^{\beta\mu}\right)\right\}
\label{2nd}
\end{flalign}
if, at the right hand side of (\ref{1st}), $\delta S_V$ is replaced with (\ref{deltSV}),  $``\!-I(\eta,V)"$ is left untouched, and yet  $``\!+I(\eta,V)"$ is integrated by-parts to involve $D_{\mu\nu}^2 = {{D}}^2 \eta_{\mu\nu} - D_{\mu}D_{\nu}-V_{\mu\nu}$ where $D_{\mu}$ is gauge-covariant derivative. This recast $\delta S_{V}$ gets to curved spacetime via (\ref{map-metric}) to become
\begin{flalign}
\delta S_{V}\left(g, \Delta^2\right) &=
 - I(g,V)\nonumber\\ &+ \int d^{4}x  \sqrt{-g} {c_V} {\mbox{Tr}}\!\left\{V^{\mu}\left( -{\mathcal{D}}_{\mu\nu}^2 + \Delta^2 g_{\mu\nu} \right)\! V^{\nu}\right\}\nonumber\\
 &+\int d^{4}x  \sqrt{-g} {c_V} {\mbox{Tr}}\!\left\{{\nabla}_{\mu} \left(g_{\alpha\beta} V^{\alpha} V^{\beta\mu}\right)\right\}
\label{3rd}
\end{flalign}
where ${\mathcal{D}}_{\mu}$ is the gauge-covariant derivative with respect to the covariant derivative $\nabla_{\mu}$ of the Levi-Civita connection ${}^{g}\Gamma^{\lambda}_{\mu\nu}$, and ${\mathcal{D}}_{\mu\nu}^2 = {\mathcal{D}}^2 g_{\mu\nu} - {\mathcal{D}}_{\mu} {\mathcal{D}}_{\nu} - V_{\mu\nu}$. 

\subsubsection{$\delta S_{V}\left(g, \Delta^2\right)$ in a New ``Curvature"}
\label{subsubsec:restore2}
Is there a simple way of killing ${{\delta S_{V}\left(g,\Delta^2\right)}}$? Yes, there is. Indeed, ${{\delta S_{V}\left(g,\Delta^2\right)}}$ vanishes identically if $\Delta^2 g_{\mu\nu}$ is replaced with  $R_{\mu\nu}\left({}^{g}\Gamma\right)$. This encouraging feature, not to be confused with derivation of gravity from self-interacting spin-2 fields in flat spacetime \cite{deser-gravity}, is pitiably problematic because $\Delta^2 g_{\mu\nu}\, {{\hookrightarrow}}\, R_{\mu\nu}\left({}^{g}\Gamma\right)$ contradicts with $\eta_{\mu\nu}\, {{\hookrightarrow}}\, g_{\mu\nu}$. If it were not for this contradiction, emergence of curvature from  $\Delta^2 g_{\mu\nu}$ would solve the CCB \cite{demir2,talk-ppc,talk-NP,talk-hepac}. 

The contradiction can be avoided by introducing, for instance,  a more general map \cite{talk-qdis}
\begin{flalign}
\label{emerge-curve-1}
\Delta^2 g_{\mu\nu}\, {{\hookrightarrow}}\, {\mathbb{R}}_{\mu\nu}\left(\Gamma\right)
\end{flalign}
in which ${\mathbb{R}}_{\mu\nu}(\Gamma)$ is the Ricci curvature of a symmetric affine connection $\Gamma^{\lambda}_{\mu\nu}$ (bearing no relationship to the Levi-Civita connection ${}^{g} \Gamma^{\lambda}_{\mu\nu}$). This new map removes  contradiction because  while ${\mathbb{R}}_{\mu\nu}\left(\Gamma\right) \leadsto \Delta^2 g_{\mu\nu}$ fixes the affine connection, $g_{\mu\nu} \leadsto \eta_{\mu\nu}$ does the metric \cite{talk-qdis}. It throws ${{\delta S_{V}\left(g,\Delta^2\right)}}$ in (\ref{3rd}) into metric-affine geometry \cite{Palatini, Palatini2} to give it the ``mass-free'' form
\begin{flalign}
{{\delta S_{V}\left(g, {\mathbb{R}}\right)}} &= - I(g,V)\nonumber\\ &+ \int d^{4}x  \sqrt{-g} {c_V} {\mbox{Tr}}\!\left\{V^{\mu}\left( -{\mathcal{D}}_{\mu\nu}^2 + {\mathbb{R}}_{\mu\nu}\left(\Gamma\right) \right)\! V^{\nu}\right\}\nonumber\\
 &+\int d^{4}x  \sqrt{-g} {c_V} {\mbox{Tr}}\!\left\{{\nabla}_{\mu} \left(g_{\alpha\beta} V^{\alpha} V^{\beta\mu}\right)\right\}
 \label{tilded-3}
 \end{flalign}
 whose by-parts integration results in
 \begin{flalign}
 {{\delta S_{V}\left(g, {\mathbb{R}}, R\right)}} = \int d^{4}x  \sqrt{-g} {c_V} {\mbox{Tr}}\!\left\{V^{\mu}\left( {\mathbb{R}}_{\mu\nu}(\Gamma) - {{R}}_{\mu\nu}(g)\right) V^{\nu}\right\}
\label{tilded-3p}
\end{flalign}
under the condition that $c_V$ must be held unchanged or, equivalently,  the EW-UV hierarchy  
\begin{flalign}
\label{fix-Uv/IR}
\frac{\Lambda_{{W}}}{\Lambda_{{U}}}\;\; {\text{must be held unchanged}}
\end{flalign}
while the affine curvature arises as in (\ref{emerge-curve-1}). This preservation  is crucial for ensuring that the SM is indeed stabilized at $\Lambda_W\ll \Lambda_U$ \cite{veltman}.

The CCB attains a solution only if ${{\delta S_{V}\left(g, {\mathbb{R}},R\right)}}$  in (\ref{tilded-3p}) is suppressed. And suppression is seen to necessitate $\Gamma^{\lambda}_{\mu\nu}$ to be close to ${}^{g}\Gamma^{\lambda}_{\mu\nu}$. Fortunately,  $\Gamma^{\lambda}_{\mu\nu}$ does really lie close to ${}^{g}\Gamma^{\lambda}_{\mu\nu}$. Indeed, as already voiced in \cite{talk-qdis} and as will be derived in equations (\ref{gamma-gammag-3}) and (\ref{expand-conn})  in Sec. \ref{subsec:deriveGR} below, $\Gamma^{\lambda}_{\mu\nu}$  assumes the form
\begin{flalign}
\label{gamma-gammag}
\Gamma^{\lambda}_{\mu\nu}={}^{g}\Gamma^{\lambda}_{\mu\nu} + {\mathcal{O}}\!\left(M_{Pl}^{-2}\right)_{\!H,V}
%+ \Delta^{\lambda}_{\mu\nu}(H,V)
%{\mathcal{O}}\left(M_{Pl} \Lambda_{{W}}^3\right) 
%{\mathcal{O}}\left(\frac{1}{M_{Pl}^2}\!\right)
\end{flalign}
as a solution to its equation of motion. It is clear that $\Gamma^{\lambda}_{\mu\nu}$ differs from ${}^{g}\Gamma^{\lambda}_{\mu\nu}$ only by Planck-suppressed ${\mathcal{O}}\!\left(M_{Pl}^{-2}\right)$ terms, which are labeled with $H,V$ to emphasize that they are made up solely of the Higgs and gauge fields. The preconditions for algebraic solutions like (\ref{gamma-gammag}) have already been revealed in \cite{Palatini, Palatini3}, and the analysis in Sec. \ref{subsec:deriveGR} will follow them. 

Under the solution (\ref{gamma-gammag}) for $\Gamma^{\lambda}_{\mu\nu}$,  the Ricci curvature takes the form
 \begin{eqnarray}
 {\mathbb{R}}_{\mu\nu}(\Gamma) = R_{\mu\nu}(g)+  {\mathcal{O}}\!\left(M_{Pl}^{-2}\right)_{\!H,V}
 \end{eqnarray}
and its replacement into the action (\ref{tilded-3p}) leads to
\begin{flalign}
{{\delta S_{V}\left(g, {\mathbb{R}},R\right)}} = \int d^{4}x  \sqrt{-g}\left\{ 0 + {\mathcal{O}}\!\left(M_{Pl}^{-2}\right)_{\!H,V} \right\}
\label{tilded-4}
\end{flalign}
as a proof of the fact that the CCB is suppressed  up to Planck-suppressed 
 ${\mathcal{O}}\!\left(M_{Pl}^{-2}\right)$ terms. It implies that color and electromagnetism  are restored and  spontaneity of the electroweak breaking is ensured modulo Planck-suppressed dimension-6 Higgs and gauge composites. 

\begin{figure}[ht]
\includegraphics[scale=0.72]{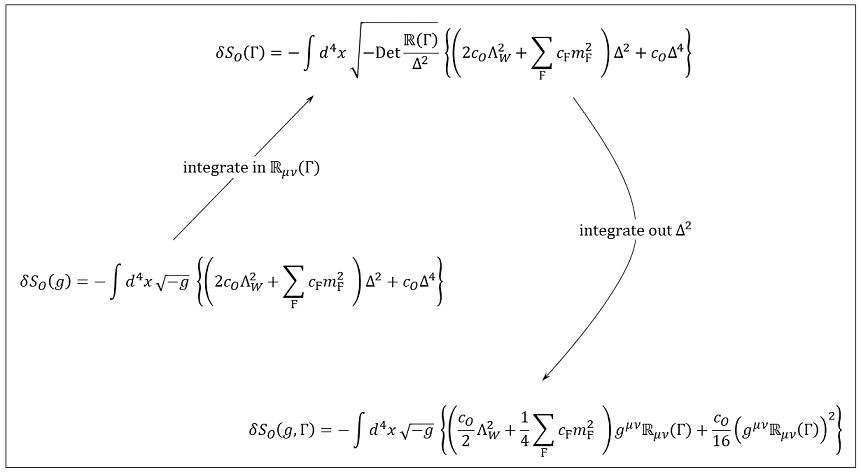}
\caption{\label{figure-reduce} 
The emergence of the affine curvature. The metrical action $\delta S_O(g)$  can be mapped into  the affine action $\delta S_O(\Gamma)$ after integrating in  ${\mathbb{R}}_{\mu\nu}(\Gamma)$. This affine action, which does not admit any matter sector due to the absence of metric, can be brought into the metric-affine form $\delta S_O(g,\Gamma)$ by integrating out $\Delta^2$. Integrating in and out both rests on the Eddington solution, which is identical to the map (\ref{emerge-curve-1}) imposed by gauge symmetry restoration.} 
\end{figure}

\subsection{How Curvature Emerges: Elasticated Flat Spacetime}
\label{subsec:elastic}
The flat spacetime is rigid. It remains flat independent of how big the energy it contains is. The quartic quantum correction $\delta S_{O}$ in (\ref{deltSO}), for instance, can no way curve it, even with infinite $\Delta$. But, flat spacetime must develop a certain degree of elasticity if gravity is to be incorporated into the effective SM. The requisite elasticity needs be generated by an extraneous mechanism as it is unlikely to arise from the effective QFT itself. One mechanism as such is Sakharov's induced gravity \cite{sakharov} in which elasticity emerges from the assumption that the quantum loops have a different metric than the tree-level action. This mechanism generates the gravitational scale as $M_{Pl}^2 \sim \Delta^2$ but leaves behind an ${\mathcal{O}}(\Delta^4)$ vacuum energy and an ${\mathcal{O}}(\Delta^2)$ Higgs mass-squared. This means that the CCP and BHP (in Table \ref{table-prob}) remain unsolved, and it is necessary to devise an alternative mechanism. 

In search for a proper mechanism, it proves useful to start with the observation that, in general, the integral 
\begin{eqnarray}
\int d^4 x \sqrt{-{\text{Det}}\left(\frac{{\mathbb{R}}\left(\Gamma\right)}{\Delta^2} \right)}
\label{vol-affine}
\end{eqnarray}
takes the form 
\begin{eqnarray}
\int d^4 x \sqrt{-g}
\label{vol-metric}
\end{eqnarray}
after integrating out the affine curvature ${\mathbb{R}}_{\mu\nu}(\Gamma)$. The reason, as was first pointed out in 
\cite{demir1}, is that the equation of motion for the affine connection 
\begin{eqnarray}
\label{eddington-eom}
{}^\Gamma \nabla_{\alpha} \left( {\text{Det}}\left(\frac{{\mathbb{R}}\left(\Gamma\right)}{\Delta^2}\right) \left( \left(\frac{{\mathbb{R}}\left(\Gamma\right)}{\Delta^2}\right)^{-1}\right)^{\mu\nu} \right) = 0,
\end{eqnarray}
which follows from (\ref{vol-affine}), is solved by 
\begin{eqnarray}
\label{eddington}
{\mathbb{R}}_{\mu\nu}\left(\Gamma\right) = \Delta^2 g_{\mu\nu}
\end{eqnarray}
with a dynamically-induced metric $g_{\mu\nu}$. This  solution, the well-known Eddington solution \cite{eddington1,eddington2}, is precisely the map (\ref{emerge-curve-1}) imposed by gauge symmetry restoration (partly because curvature scale in (\ref{vol-affine}) is set to $\Delta^2$).

The meaning of the relation (\ref{eddington}) is that curvature emerges as elastication of the rigid flat 
spacetime. To see how this happens, it proves efficacious to focus on the colossal vacuum energy in (\ref{deltSO}). 
The metrical action $\delta S_O(g)$, obtained from $\delta S_O(\eta)$ in (\ref{deltSO}) via (\ref{map-metric}), can be mapped into the affine action $\delta S_O(\Gamma)$ by integrating in  ${\mathbb{R}}_{\mu\nu}\left(\Gamma\right)$ via the Eddington solution (\ref{eddington}). This ``integrating in" stage,   illustrated in  Fig. \ref{figure-reduce}, represents the elastication of the flat spcetime \cite{demir1}.  The problem here is that the affine geometry of $\delta S_O(\Gamma)$ is devoid of any metrical structure to accommodate  $\delta S_{log}$ and $\delta S_{V}$. The remedy is to integrate out $\Delta^2$  to get the metric-affine action $\delta S_O(g,\Gamma)$ in Fig. \ref{figure-reduce}. The metric-affine geometry  accommodates all parts of the SM effective action and the resulting framework, which has curvature incorporated into the effective SM, is identical to that obtained via (\ref{map-metric}) in Sec. \ref{subsec:notrestore} followed by (\ref{emerge-curve-1}) in Sec. \ref{subsubsec:restore2}. This means that the maps (\ref{map-metric}) and (\ref{emerge-curve-1}) have a dynamical origin and serve as equivalence relations for incorporating curvature into effective QFTs. 

The emergence mechanism here  is similar, at least in philosophy, to that of Sakharov \cite{sakharov} in that curvature is tied to quantum effects. It is based on emergence of the curvature not that of the spacetime itself, and differs thus from mechanisms like  the entropic gravity of Verlinde \cite{emerge-verlinde} and entanglement gravity of Raamsdonk \cite{emerge-raamsdonk}.

\subsection{Symmergence Principle}
\label{sec-symmergence}
The results of Sec. \ref{subsubsec:restore2} and Sec. \ref{subsec:elastic} above can be organized as a principle to apply to UV sensitivies of any given QFT. Indeed, it turns out that gravity can be incorporated into a flat spacetime QFT with UV-IR gap $\Delta$ and  metric $\eta_{\mu\nu}$ by first letting \cite{eqp}
\begin{flalign}
\eta_{\mu\nu}\, {{\hookrightarrow}}\, g_{\mu\nu}
\end{flalign}
and then \cite{talk-qdis}
\begin{flalign}
\Delta^2 g_{\mu\nu}\, {{\hookrightarrow}}\, {\mathbb{R}}_{\mu\nu}\left(\Gamma\right)
\end{flalign}
so that if 
\begin{flalign}
\label{reduce}
 {\mathbb{R}}_{\mu\nu}\left(\Gamma\right) \leadsto {{R}}_{\mu\nu}\left({}^{g}\Gamma\right)
\end{flalign}
then CCB gets suppressed and GR emerges as a gauge {\underline{symm}}etry restoring em{\underline{ergent}} gravity or, briefly, {\it symmergent gravity}. This three-stage mainframe will define what shall be called {\it symmergence} in what follows. 

\section{Incorporation of GR and the Requisite NP}
Having set up the symmergence principle, it is now time to determine under what conditions GR arises correctly. It will be found below that it cannot arise without a nontrivial NP sector. 

\subsection{Metric-Affine Gravity}
Affine gravity, as in $\delta S_O(\Gamma)$ in Fig. \ref{figure-reduce}, involves only the affine connection \cite{eddington1,eddington2}. Metrical gravity, the GR itself, is based solely on the metric tensor. The metric-affine gravity (MAG), on the other hand, comprises, as in $\delta S_O(g,\Gamma)$ in Fig. \ref{figure-reduce}, both metric and affine connection as two independent dynamical variables \cite{Palatini,Palatini2,Palatini3}. The MAG reduces to the so-called Palatini formalism \cite{Palatini} if connection doe not appear in the matter sector. It is equivalent to the GR. (In general, affine connection spreads to matter sector via at least the spin connection.) 

MAG may or may not lead to GR. It depends on couplings of the affine connection. Indeed, solving the equation of motion for affine connection to integrate it out leads to the Levi-Civita connection of the metric plus possible contributions from other fields, including the affine curvature \cite{Palatini3,romano}. Then, reduction to the GR can put constraints even on the matter sector. The sections below, starting with Sec. \ref{subsec:deriveGR}, will be devoted to the GR limit of the MAG arising from the SM effective action.

\subsection{Necessity of NP}
\label{subsec:necessity}
Under (\ref{map-metric}) and (\ref{emerge-curve-1}) or under the  emergence mechanism in Fig. \ref{figure-reduce}, the Higgs and vacuum sectors of the flat spacetime effective action in (\ref{action-flat}) transform as
\begin{flalign}
\delta S_{O} + \delta S_{H} \hookrightarrow \int d^4x \sqrt{-g} \left\{- q^{\mu\nu} {\mathbb{R}}_{\mu\nu}(\Gamma) + \frac{c_O}{16} \left(g^{\mu\nu} {\mathbb{R}}_{\mu\nu}(\Gamma)\right)^2 \right\}
\label{action-affine-2}
\end{flalign}
after defining
%the scalar affine curvature
%\begin{eqnarray}
%\label{scalar-affine}
%{\mathbb{R}}(g,\Gamma)=g^{\mu\nu} %{\mathbb{R}}_{\mu\nu}(\Gamma)
%\end{eqnarray}
%and introducing
\begin{flalign}
q_{\mu\nu} = \left(\frac{c_O}{2} \Lambda_{{W}}^2 + \frac{1}{4}\sum_{\digamma}c_\digamma m_\digamma^2 + \frac{c_H}{4} H^{\dagger} H + \frac{c_O}{8} g^{\alpha\beta} {\mathbb{R}}_{\alpha\beta}(\Gamma)\right) g_{\mu\nu}
\label{qmunu-SM}
\end{flalign}
for convenience. The right-hand side of (\ref{action-affine-2}) forms the curvature sector of the SM effective action in metric-affine geometry \cite{Palatini,Palatini2,Palatini3}. The apparent gravitational scale, $\frac{c_O}{2} \Lambda_{{W}}^2 + \frac{1}{4}\sum_{\digamma}c_\digamma m_\digamma^2$, is wrong in  both sign ($c_O < 0$ in the SM) and size ($\sum_{\digamma} c_\digamma m_\digamma^2 < \Lambda_{{W}}^2$ in the SM). It is for this reason that an NP sector (made up of entirely new fields $\digamma^{\prime}$) must be introduced so that the new gravitational scale
\begin{flalign}
\label{Mpl}
\left(c_O + c_{O^{\prime}}\right)  \Lambda_{{W}}^2 + \frac{1}{2}\sum_{f=\digamma,\digamma^{\prime}} c_f m_f^2 
\end{flalign}
can come out right thanks to either the NP spectrum ($c_{O^{\prime}}$) or the NP scale ($m_{\digamma^{\prime}}$). In fact, its one-loop value
\begin{flalign}
\label{Mpl-1loop}
\frac{1}{64 \pi^2}\left({\mbox{Str}}\left[1\right] \Lambda_{{W}}^2 + {\mbox{Str}}\left[m^2\right]\right)
\end{flalign}
makes it clear that the NP must have either a crowded (${\mbox{Str}}\left[1\right]\sim 10^{35}$) \cite{talk-NP,talk-hepac,talk-qdis} or a heavy (${\mbox{Str}}\left[m^2\right] \gtrsim M_{Pl}^2$)  bosonic sector.

\subsection{Incorporation of GR}
\label{subsec:deriveGR}
The problem is to determine how MAG can reduce to GR. This reduction is decided by the dynamics of the affine connection. And part of the total SM + NP action that governs the dynamics 
\begin{flalign}
 \int d^4x \sqrt{-g}\, \left\{-Q^{\mu\nu} {\mathbb{R}}_{\mu\nu}(\Gamma) + \frac{1}{16}(c_O + c_{O^{\prime}}) \left(g^{\mu\nu} {\mathbb{R}}_{\mu\nu}(\Gamma)\right)^2\right\}
\label{action-affine-2px}
\end{flalign}
involves 
\begin{flalign}
\label{q-tensor}
Q_{\mu\nu} &= \left(\frac{1}{2}(c_O + c_{O^{\prime}}) \Lambda_{{W}}^2 + \frac{1}{4}\sum\limits_{f=\digamma,\digamma^{\prime}} c_f m_f^2 +  \frac{1}{4} \sum\limits_{\phi=H,H^{\prime}}c_\phi \phi^{\dagger} \phi + \frac{1}{8}(c_O + c_{O^{\prime}}) g^{\alpha\beta} {\mathbb{R}}_{\alpha\beta}(\Gamma)\right) g_{\mu\nu}\nonumber\\
&- \sum\limits_{{\mathcal{V}}=V,V^{\prime}} {c_{\mathcal{V}}} {\mbox{Tr}}\left\{{\mathcal{V}}_{\mu}{\mathcal{V}}_{\nu}\right\}
\end{flalign}
as an extension of $q_{\mu\nu}$ in (\ref{qmunu-SM}) to the gauge and NP sectors.  It gathers affine curvature terms from (\ref{tilded-3p}) (after extending it with the NP gauge bosons $V^{\prime}_{\mu}$ with loop factors $c_{V^{\prime}}$) and (\ref{action-affine-2}) (after augmenting it with the  NP vacuum energy with the loop factor $c_{O^{\prime}}$ and all of the NP fields with loop factors $c_{\digamma^{\prime}}$). 
%It is clear that $q^{\prime}_{\mu\nu} =  q_{\mu\nu}\big(c_O \rightarrow c_{O^{\prime}},$ $c_H \rightarrow c_{H^{\prime}},$ $c_\digamma \rightarrow c_{\digamma^{\prime}} \big)$.

The equation of motion for $\Gamma^{\lambda}_{\mu\nu}$, stipulated  by stationarity of the action (\ref{action-affine-2px}) against variations in $\Gamma^{\lambda}_{\mu\nu}$, turns out to be a metricity condition on $Q_{\mu\nu}$
\begin{flalign}
\label{gamma-eom}
{}^{\Gamma}\nabla_{\lambda} Q_{\mu\nu} = 0
\end{flalign}
and  possesses the generic solution \cite{Palatini,Palatini3}
\begin{flalign}
\label{gamma-gammag-2}
\Gamma^{\lambda}_{\mu\nu}={}^{g}\Gamma^{\lambda}_{\mu\nu} + \frac{1}{2} (Q^{-1})^{\lambda\rho} \left({{\nabla}}_{\mu} Q_{\nu\rho} + {{\nabla}}_{\nu} Q_{\rho\mu} - {{\nabla}}_{\rho} Q_{\mu\nu}\right)
\end{flalign}
which expresses $\Gamma^{\lambda}_{\mu\nu}$ in terms of its own curvature ${\mathbb{R}}_{\mu\nu}(\Gamma)$,  as evidenced by $Q_{\mu\nu}$ itself.  This dependence on curvature is critical because if it is the case then (\ref{gamma-gammag-2}) acts as a differential equation for  $\Gamma^{\lambda}_{\mu\nu}$, and its solution carries extra geometrical degrees of freedom not found in the Levi-Civita connection ${}^{g}\Gamma^{\lambda}_{\mu\nu}$ \cite{Palatini3}. This means that ${\mathbb{R}}_{\mu\nu}(\Gamma)$ must disappear from $Q_{\mu\nu}$ for GR to be able to symmerge, and it does so  if 
\begin{flalign}
c_O + c_{O^{\prime}} = 0\;\; \text{or, equivalently,}\;\; {\mbox{Str}}\left[1\right] = 0
\label{f-b-b}
\end{flalign}
or, equivalently, SM + NP has equal bosonic and fermionic degrees of freedom. This fermi-bose balance does of course not mean a proper SUSY model since SM-NP couplings do not have to support a SUSY structure \cite{beyond}. All it does is to trim $Q_{\mu\nu}$ in (\ref{q-tensor}) down to
\begin{flalign}
\label{q-tensor-0}
{\mathring{Q}}_{\mu\nu} = \left(\frac{{{M}}^2}{2} + \frac{1}{4} \sum\limits_{\phi=H,H^{\prime}}c_\phi \phi^{\dagger} \phi\right) g_{\mu\nu}  - \sum\limits_{{\mathcal{V}}=V,V^{\prime}} {c_{\mathcal{V}}} {\mbox{Tr}}\left\{{\mathcal{V}}_{\mu}{\mathcal{V}}_{\nu}\right\}
\end{flalign}
in which
\begin{flalign}
\label{Mpl-2}
{{M}}^2 = \sum\limits_{f=\digamma,\digamma^{\prime}} c_f m_f^2 \xlongequal{\text{1-loop}} \frac{{\mbox{Str}}\left[m^2\right]}{64 \pi^2}
\end{flalign}
 is the apparent gravitational scale. Now, GR can symmerge because $\Gamma^{\lambda}_{\mu\nu}$  is an auxiliary field (algebraic solution) having no geometrical content beyond ${}^{g}\Gamma^{\lambda}_{\mu\nu}$ \cite{Palatini,Palatini3}. Indeed, after (\ref{q-tensor-0}),  the affine connection in (\ref{gamma-gammag-2}) takes the form
 \begin{flalign}
\label{gamma-gammag-3}
{{\Gamma}}^{\lambda}_{\mu\nu}={}^{g}\Gamma^{\lambda}_{\mu\nu} + \frac{1}{2} ({\mathring{Q}}^{-1})^{\lambda\rho} \left({{\nabla}}_{\mu} {\mathring{Q}}_{\nu\rho} + {{\nabla}}_{\nu} {\mathring{Q}}_{\rho\mu} - {{\nabla}}_{\rho} {\mathring{Q}}_{\mu\nu}\right)
\end{flalign}
 to contain only the Levi-Civita connection and the scalars $\phi$ and vectors ${\mathcal{V}}_{\mu}$ in ${\mathring{Q}}_{\mu\nu}$. Given the enormity of $M$,  this expression can always be expanded order by order in $1/M^2$ to get 
\begin{flalign}
{{\Gamma}}^{\lambda}_{\mu\nu} = {}^{g}\Gamma^{\lambda}_{\mu\nu}  + \frac{1}{M^2}\left(\nabla_{\mu}{\mathring{Q}}_{\nu}^{\lambda} + \nabla_{\mu}{\mathring{Q}}_{\nu}^{\lambda} - \nabla^{\lambda}{\mathring{Q}}_{\mu\nu}\right) +  {\mathcal{O}}\!\left(M^{-4}\right)_{\phi,{\mathcal{V}}}
\label{expand-conn}
\end{flalign}
using the notation in (\ref{gamma-gammag}) for the remainder of the expansion. The higher-order ${\mathcal{O}}\!\left(M^{-4}\right)_{\phi,{\mathcal{V}}}$ terms are straightforwardly computed from the exact expression (\ref{gamma-gammag-3}) order by order in $1/M^2$. In accordance with (\ref{expand-conn}), the affine curvature expands as 
\begin{flalign}
{\mathbb{R}}_{\mu\nu}(\Gamma) = R_{\mu\nu}({}^{g}\Gamma) + \frac{1}{M^2} \left(\nabla^2\right)_{\mu\nu}^{\alpha\beta} {\mathring{Q}}_{\alpha\beta} +  {\mathcal{O}}\!\left(M^{-4}\right)_{\phi,{\mathcal{V}}}
\label{expand-curv}
\end{flalign}
where $\left(\nabla^2\right)_{\mu\nu}^{\alpha\beta} = \nabla^{\alpha}\, \nabla_{\mu} \delta^{\beta}_{\nu} + \nabla^{\alpha}\, \nabla_{\nu} \delta^{\beta}_{\mu} - \Box \delta^{\alpha}_{\mu} \delta^{\beta}_{\nu} - \nabla_{\nu}\, \nabla_{\mu} g^{\alpha\beta}$ \cite{grav-prop}. This expansion leads to the reductions 
\begin{flalign}
\int d^4x \sqrt{-g} \frac{{{M}}^2}{2}g^{\mu\nu} {\mathbb{R}}_{\mu\nu}(\Gamma) &\leadsto  \int d^4x \sqrt{-g} \left\{\frac{{{M}}^2}{2}  R(g) + {\mathcal{O}}\!\left(M^{-2}\right)_{\phi,{\mathcal{V}}} \right\}
\end{flalign}
\begin{flalign}
\int d^4x \sqrt{-g} \sum\limits_{\phi=H,H^{\prime}} \phi^{\dagger} \phi g^{\mu\nu} {\mathbb{R}}_{\mu\nu}(\Gamma) &\leadsto \int d^4x \sqrt{-g} \left\{ \sum\limits_{\phi=H,H^{\prime}} \phi^{\dagger} \phi R(g) + {\mathcal{O}}\!\left(M^{-2}\right)_{\phi,{\mathcal{V}}}\right\}
\end{flalign}
\begin{flalign}
\int d^4x \sqrt{-g} \sum\limits_{{\mathcal{V}}=V,V^{\prime}} {\mbox{Tr}}\!\Big[{\mathcal{V}}^{\mu}\left({\mathbb{R}}_{\mu\nu}(\Gamma)-R_{\mu\nu}(g)\right) {\mathcal{V}}^{\nu}\Big] &\leadsto \int d^4x \sqrt{-g} \left\{0 + {\mathcal{O}}\!\left(M^{-2}\right)_{\phi,{\mathcal{V}}}\right\}
\label{gauge-reduce}
\end{flalign}
so that the SM + NP affine action in (\ref{action-affine-2px}) reduces to 
\begin{flalign}
 \int d^4x \sqrt{-g}\, \left\{-\left(\frac{{{M}}^2}{2} + \frac{1}{4} \sum\limits_{\phi=H,H^{\prime}}
 c_\phi \phi^{\dagger} \phi\right)R(g) + {\mathcal{O}}\!\left(M^{-2}\right)_{\phi,{\mathcal{V}}} \right\}
\label{action-affine-2pxy}
\end{flalign}
to yield the Einstein-Hilbert action for non-minimally coupled scalars $\phi=H, H^{\prime}$ \cite{non-min-curve}. Here,  ${\mathcal{O}}\!\left(M^{-2}\right)_{\phi,{\mathcal{V}}}$ terms can be constructed from the exact solution (\ref{gamma-gammag-3}) order by order in $1/M^2$. Unlike the generic effective field-theoretic approach \cite{weinberg-eff}, the remainder  ${\mathcal{O}}\!\left(M^{-2}\right)_{\phi,{\mathcal{V}}}$ assumes a specific structure coordinated by (\ref{gamma-gammag-3}) through ${\mathring{Q}}_{\mu\nu}$ in (\ref{q-tensor-0}).

The GR gets properly incorporated if Planck scale is generated correctly. In view of (\ref{action-affine-2pxy}), it is given by 
\begin{flalign}
\label{Mpl2}
M_{Pl}^2 &= {{M}}^2 + \frac{1}{2} \sum\limits_{\phi=H,H^{\prime}} {c_\phi} \langle \phi^{\dagger} \phi\rangle \xlongequal{\text{1-loop}} \frac{{\mbox{Str}}\left[m^2\right]}{64 \pi^2}  +  \frac{1}{2} \sum\limits_{\phi=H,H^{\prime}} {c_\phi^{(1)}} \langle \phi^{\dagger} \phi\rangle
\end{flalign}
in which vacuum contribution becomes significant when $\sum_{H^{\prime}} c_{H^{\prime}} \langle H^{\prime\dagger}H^{\prime}\rangle \sim M^2$. The first part, identical to (\ref{Mpl-2}), implies that the bosonic sector of the NP must be either crowded ($n_{b^{\prime}}\sim 10^{35}$ and $m_{\digamma^{\prime}}\gtrsim \Lambda_W$) \cite{demir2,talk-NP,talk-hepac,talk-qdis} or heavy enough ($n_{b^{\prime}}\sim$ much less yet $m_{\digamma^{\prime}}\lesssim \Lambda_U$) for it to be able to induce the trans-Planckian scale $8 \pi M_{Pl}$. 

The NP sector, needed for induction of the gravitational scale as in (\ref{Mpl}) or  (\ref{Mpl2}), can have nontrivial structure. Indeed, given the fermi-bose balance of SM+NP and given also that ${\mbox{Str}}\left[m^2\right]$ in  (\ref{Mpl2}) is reminiscent of the mass sum rule in broken SUSY \cite{susy,beyond}, it becomes apparent that the SM and NP, with the seesawic couplings in (\ref{sm-NP-coup}), may well be the remnants of a trans-Planckian SUSY broken around $8 \pi M_{Pl}$. Pivotally, this can provide an explanation for  why  symmergence must start with flat spacetime SM + NP in that the alleged SUSY theory, whose breaking occurs with an anomalous $U(1)$ factor \cite{susy} to generate the nonzero sum rule in (\ref{Mpl2}), cannot couple to gravity due to the anomaly. In other words, flat spacetime QFT can be taken as a signature of SUSY. These features can be useful for probing trans-Planckian physics but a complete NP model, which presumably is subject matter of a whole different study, is not an urgency as far as the gravitational constant is concerned.

\section{Observational and Experimental Tests}
This section collects various predictions and experimentally testable features from both the gravity and NP sectors. They can play a crucial role in revealing and testing the physics of symmergence.

\subsection{CCB is suppressed}
Nullification of the gauge part in  (\ref{gauge-reduce}) ensures that color and electromagnetism are restored and electroweak breaking is rendered spontaneous up to doubly Planck-suppressed terms involving  $\phi$ and ${\mathcal{V}}_{\mu}$. This is important for assuring neutralization of the CCB in accordance with the symmergence principle. To that end, exact form of the affine connection in (\ref{gamma-gammag-3}), which is never derivative-free, ensures that gauge boson masses are impossible to arise at any order in $1/M^2$. Indeed, iteration of  (\ref{action-affine-2pxy}) up to next order 
\begin{flalign}
 \int d^4x \sqrt{-g}\, \left\{-\left(\frac{{{M}}^2}{2} + \frac{1}{4} \sum\limits_{\phi=H,H^{\prime}}
 c_\phi \phi^{\dagger} \phi\right)R(g) + \frac{1}{M^2}{\mathring{Q}}^{\mu\nu} \left(\nabla^2\right)_{\mu\nu}^{\alpha\beta} {\mathring{Q}}_{\alpha\beta} +  {\mathcal{O}}\!\left(M^{-4}\right)_{\phi,{\mathcal{V}}} \right\}
\label{high-order}
\end{flalign}
exemplifies that what arise at each order involve  derivatives of $\phi$ and ${\mathcal{V}}_{\mu}$.

\subsection{The UV boundary might have a say in CCP}
The total vacuum energy at one loop
\begin{flalign}
V\left(\Lambda_U\right) = V_{tree}\left(\langle\phi\rangle\right) + \frac{1}{64
\pi^2}{\rm Str}\!\left[{m^{4}}\left(\log \frac{m^{2}}{\Lambda_{{U}}^2} - \frac{1}{2}\right) \right]
\label{VV}
\end{flalign}
is composed of the tree-level potential plus the loop corrections involving the UV scale $\Lambda_U$. It does not have to be small. In fact, it can readily exceed the current observational bounds \cite{lam-cdm} to give cause to the CCP \cite{ccp}. Needless to say, the CCP arises from not the $\Delta^4$ sensitivity in (\ref{deltSV}) (which is neutralized by symmergence as in  (\ref{action-affine-2})) but the energies released by the QCD, electroweak and possible NP transitions as well as the zero-point energies of all the SM + NP fields. The question of how these distinct energy sources, making up (\ref{VV}), can conspire to yield the observed value $V_{obs} = \left(2.57\times 10^{-3}\ {\rm eV}\right)^4$ \cite{lam-cdm} is what the CCP is all about \cite{ccp}. Indeed, the empirical equality
\begin{eqnarray}
V\left(\Lambda_U^0\right) = V_{obs}
\label{fix-LambdaU}
\end{eqnarray}
gives a fix on the UV boundary in that the solution $\Lambda_U = \Lambda_U^0$ expresses $\Lambda_U$ in terms of the SM and NP parameters. This solution can hardly make any sense unless the aforementioned energy sources are put in relation by some governing rule. Indeed, a correlation rule, which might be a remnant of the trans-Planckian SUSY above $8\pi M_{Pl}$ \cite{susy-ccp}, can prevent fine-tuning and render $\Lambda_U^0$ physical.  

Symmergent gravity and the NP it necessitates agree with all the existing bounds thanks to their prediction that gravity is Einstein (as in \cite{lam-cdm}), missing matter can be ebony (as in \cite{cdm-current}), and non-SM interactions are not a necessity (as in \cite{exotica}). In case of  experimental discoveries, however, they can be probed with conclusive tests via the dark matter and SM-NP couplings.

\subsection{Higher-curvature terms are absent}
Symmergence leads uniquely to Einstein gravity. It excludes all higher-curvature terms. This is an important result since extinction of higher-curvature terms is impossible to guarantee in a theory in which general covariance is the only symmetry. To this end, the question of why it is not $f(R)$ gravity \cite{fR} but just the GR has an answer  in symmergence. And current observations \cite{lam-cdm} seem to have already confirmed the symmergence. 

\subsection{Higgs-curvature coupling is predicted}
The non-minimal Higgs-curvature coupling \cite{non-min-curve}, symmergement of the Higgs quadratic UV sensitivity in (\ref{deltSH}), is a model-dependent loop factor. It has the numerical value
\begin{eqnarray}
\frac{c_H}{4} \approx 1.29\times 10^{-2} 
\label{prediction}
\end{eqnarray}
as follows from (\ref{veltman-factor}). This prediction is specific to the SM spectrum. It  changes by an amount $\propto \lambda^2_{H\digamma^{\prime}}$ in the presence of the NP. This change, as ensured by the bound (\ref{sm-NP-coup}), can be significant only if the NP lies close to the electroweak scale. This means that the questions of if the underlying mechanism is symmergence and if the NP is heavy or
not can be conclusively probed by measuring $c_H$. 

Obviously, $c_H$ is too small to facilitate the Higgs inflation \cite{Higgs-inflation}. It can, nevertheless, give cause to observable effects in strong gravitational fields ({\it e.g.} early universe and black hole horizons) or Planckian-energy particle scatterings \cite{non-min-cinli}. 

The non-minimal $H^{\prime}$-curvature coupling \cite{non-min-curve} is expected to have similar features. Indeed,  $c_{H^{\prime}}$ in (\ref{action-affine-2pxy}) can be directly computed with a concrete NP model. It may differ significantly from $c_H$. In fact,  what drives cosmic inflation could well be $H^{\prime}$ not $H$ \cite{demir2,talk-NP}.

\subsection{Couplings run nearly as in the SM}
Symmergence leaves behind only logarithmic sensitivity to the UV boundary. 
This remnant sensitivity, with all gauge symmetries restored, can naturally be interpreted in the language of dimensional regularization. Indeed, the formal correspondence \cite{cutoff-dimreg,demir2,talk-NP,talk-qdis}
\begin{eqnarray}
\log\left(\frac{\Lambda^{0}_{U}}{\Lambda_{{W}}}\right)^2 = \frac{1}{\epsilon} + \log\left(\frac{\Lambda_U^{0}}{\mu}\right)^2
\end{eqnarray} 
expresses all amplitudes in terms of the renormalization scale $\mu$ after subtracting $1/\epsilon$ terms in ${\overline{\text{MS}}}$ scheme. Their independence from $\mu$ leads to the usual RGEs. The important point here is that  dimensional regularization arises as a result, not as an arbitrarily chosen regularization method.  The testable feature is that the SM couplings and masses must run as in the SM at all scales unless the NP lies close to $\Lambda_W$. It can be directly tested at present \cite{coupling-LHC} and future \cite{coupling-FCC,coupling-ILC} colliders. 

\subsection{SM-NP coupling is not a necessity}
Emergence of the gravitational scale  does not necessitate any SM-NP coupling. The SM and NP do not have to interact. The NP sector can therefore come in three different kinds: 

\begin{figure}[ht]
\centering
\includegraphics[scale=0.37]{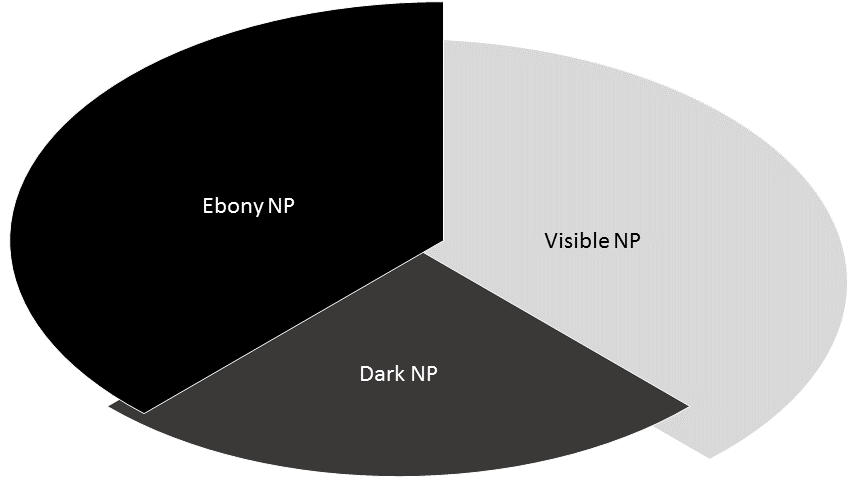}
\caption{The three kinds of the NP sectors.}
\label{figure-NP}
\end{figure}

\begin{enumerate}
    \item {\bf Ebony NP.} This kind of NP has no non-gravitatinal couplings to the SM \cite{ebony}. It forms an insular sector (composed of, for instance, high-rank non-Abelian gauge fields and fermions \cite{demir2,talk-NP}). In view of the present searches, which seem all negative, this ebony, pitch-dark NP shows good agreement with all the  available data \cite{cdm-current}.
    \item {\bf Dark NP.} This type of NP is neutral under the SM but couples to it via Higgs, hypercharge  and lepton portals \cite{demir2,portals,neutrino-portal}. Indeed, at the renormalizable level, scalars $H^{\prime}$, Abelian vectors $V^{\prime}_{\mu}$ and fermions $N^{\prime}$ can  couple to the SM via  
     \begin{flalign}
 S_{int} =\!\!\int d^4x \sqrt{-g} \Big\{\! \lambda^2_{HH^{\prime}} \left(H^{\dagger} H\right) \left(H^{\prime\dagger} H^{\prime}\right) + \lambda_{B Z^{\prime}} B_{\mu\nu} Z^{\prime\mu\nu} + \left[\lambda_{HN^{\prime}}
\overline{L} H N^{\prime} + {\mbox{h.c.}} \right]\! \Big\}
\label{action-sm-NP-int}
    \end{flalign}
    in which the couplings $\lambda^2_{HH^{\prime}}$, $\lambda_{B Z^{\prime}}$, $\lambda_{HN^{\prime}}$ can take, none to significant, a wide range of values with characteristic experimental signals \cite{cdm-current, dm-LHC}.
    
    \item {\bf Visible NP.} In this case, the NP is partly or wholly charged under the SM ({\it e.g.}, $H^{\prime}$ in (\ref{action-sm-NP-int}) can be an SU(2) doublet) \cite{demir2}. It couples to the SM with SM coupling strengths, and it seems to have already  been sidelined by the current bounds \cite{exotica,visible}. 
\end{enumerate}
The NP sector, whose subsectors are depicted in Fig. \ref{figure-NP}, can be of any composition (ebony to dark), can have any couplings (none to significant) and can lie anywhere (from $\Lambda_{{W}}$ to $\Lambda_{{U}}$) as long as the gravitational scale in (\ref{Mpl}) comes out right. Non-necessity of any sizable NP-SM couplings is what distinguishes the NP  of the symmergence from SUSY, extra dimensions, composites models and others \cite{beyond}
where a sizable SM-NP coupling is a necessity. It is this sizable coupling of theirs that make such NP models sidelined under the LHC bounds \cite{exotica}. 

\subsection{Electroweak stability restricts SM-NP coupling}
The Higgs part of (\ref{action-affine-2}) 
makes it clear that the quadratic UV sensitvity in flat spacetime changes to Higgs-curvature coupling in metric-affine spacetime. This solves the BHP \cite{veltman}. In the presence of NP, however, the Higgs sector is destabilized by not only the BHP but also the SM-NP coupling \cite{lhp2,demir2,talk-NP}. Indeed, the SM-NP interactions in (\ref{action-sm-NP-int}), for instance, lead to a new Higgs mass correction (in addition to $\delta S_{H}$ in (\ref{deltSH})) 
\begin{flalign}
{\delta S_{H}}^{\prime} &=\int d^4x \sqrt{-g} \sum\limits_{\digamma^{\prime}} {\tilde{c}}_H \lambda^2_{\digamma \digamma^{\prime}}  {m^2_{ \digamma^{\prime}}}\log\frac{{m_{\digamma^{\prime}}}}{\Lambda_{{U}}}\, H^{\dagger} H
\label{sm-NP-higgs}
\end{flalign}

\begin{figure}[ht]
\centering
\includegraphics[scale=0.57]{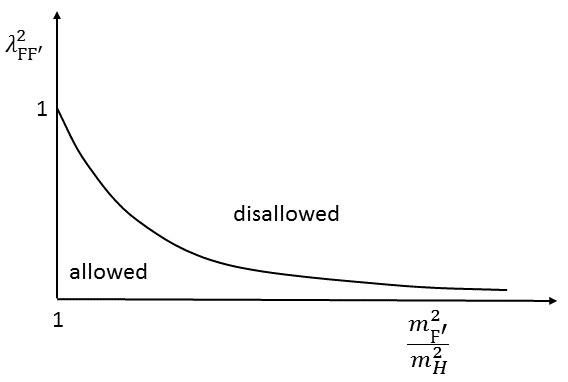}
\caption{The allowed and disallowed regions according to the electroweak stability bound in (\ref{sm-NP-coup}).}
\label{figure2}
\end{figure}

with a loop factor ${\tilde{c}}_H$. This is sensitive to not ${\Lambda_{{U}}}$ but ${{m_{\digamma^{\prime}}}}$. The problem is that heavier the NP larger the shift in the Higgs mass and stronger the destabilization of the electroweak scale. The more serious problem is that symmergence can do nothing about it. In fact, no UV completion can do anything about it. The reason is that problem is caused by coupling between two scale-split QFTs and there can hardly exist any solution other than suppression of the coupling itself. In SUSY, extra dimensions and compositeness  SM-NP coupling is a must for them to work \cite{beyond,lhp2}. In symmergence, however, SM-NP coupling is not a necessity \cite{demir2,talk-NP,talk-hepac,talk-qdis}. It works with any coupling strengths such as the seesawic ones
\begin{flalign}
\label{sm-NP-coup}
{\lambda^2_{\digamma \digamma^{\prime}}  \lesssim \frac{m_H^2}{{m^2_{ \digamma^{\prime}}}}}
\end{flalign}
with which the Higgs mass shift in (\ref{sm-NP-higgs}) falls below the Higgs mass to ensure stability of the electroweak scale. This seesawic structure, explicated in Fig. \ref{figure2}, implies that heavier the NP weaker its coupling to the SM.

\begin{figure}[ht]
\includegraphics[scale=1.0]{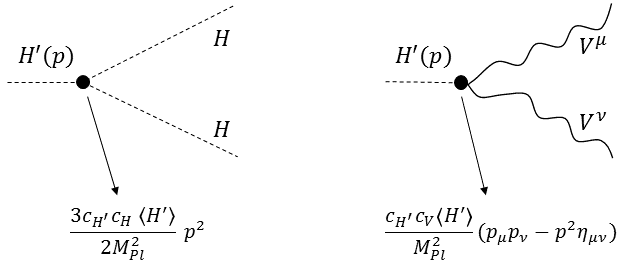}
\caption{\label{figure-scatt} The NP scalar $H^{\prime}$ can have sizeable decay rates at large VEV.}
\end{figure}

\subsection{The NP can be probed in cosmic rays}
The Planck-suppressed $\phi$ and ${\mathcal{V}}_{\mu}$ composites in (\ref{high-order}) do not produce any mass terms. They can produce, however, sizable signals if the NP scalars $H^{\prime}$ develop large VEVs (compared to $M_{Pl}$). This effect, as depicted in Fig. \ref{figure-scatt}, leads to observable effects. The $H^{\prime}$ decays  into two gluons of GZK energy \cite{GZK}, for instance, occurs at a rate
\begin{flalign}
\Gamma\left(g g\right) \approx \left(\frac{c_{H^{\prime}}}{10^{-2}}\right)^2 \left(\frac{c_{g}}{10^{-2}}\right)^2 \left(\frac{m_{H^{\prime}}}{10^{11}\ {\rm GeV}}\right)^5 \left(8.08\ {\rm min}\right)^{-1}
\end{flalign}
with $\langle H^{\prime}\rangle \simeq m_{H^{\prime}}$.
These gluons can partake in ultra high energy cosmic rays \cite{cosmic-ray} upon hadronization. The diphotons, produced similarly, can contribute to diffuse gamma-ray background \cite{gamma-ray}. These events, which weaken at low $\langle H^{\prime}\rangle$, are cosmic probes of symmergence.

\subsection{The NP can be probed in colliders}
One immediate implication of the bound (\ref{sm-NP-coup}) is that heavier the NP larger the luminosity needed to probe them at colliders. This can prove important in designing accelerators and detectors, as implied by the exemplary analyses below.

\begin{figure}[ht]
\includegraphics[scale=0.88]{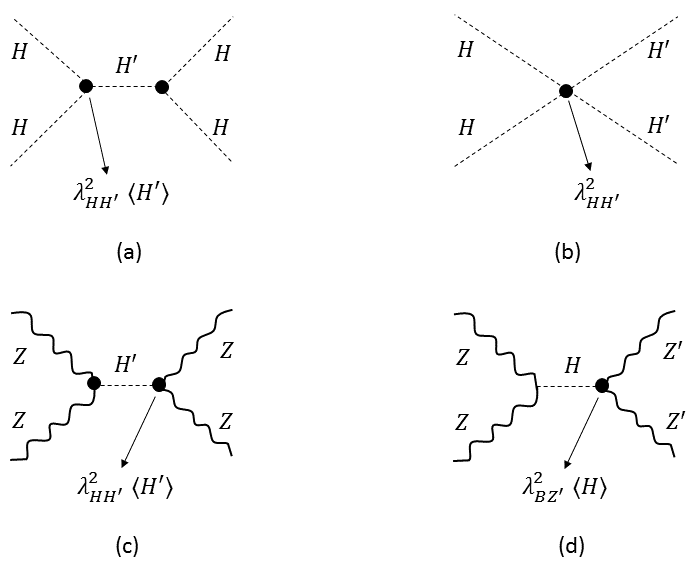}
\caption{Basic search channels for NP scalars ((a) and (b)) and NP vectors ((c) and (d)).}
\label{figure-feyn}
\end{figure}

\subsubsection{NP Scalars and Vectors}
One way to see if the underlying model is symmergence or not is to measure scalar and vector masses and then determine if their production cross sections comply with the seesawic couplings in (\ref{sm-NP-coup}). For instance, $m_{H^{\prime}}$ can be measured from either of  $H H\rightarrow H^{\prime} \rightarrow H H$ in Fig. \ref{figure-feyn} (a) or $H H\rightarrow H^{\prime} H^{\prime}$ in Fig. \ref{figure-feyn} (b) such that the ratio of their cross sections
\begin{eqnarray}
\frac{\sigma\left({{H}}H\rightarrow H^{\prime}\rightarrow {{H}}H  \right)}{\sigma\left({{H}}H\rightarrow H^{\prime}H^{\prime}\right)} \simeq \left(\frac{m^2_H}{m^2_{H^{\prime}}} \right)^2
\label{scatter2}
\end{eqnarray}
acts as an efficient probe of symmergence. This procedure works also for  the $Z^{\prime}$ scatterings in Fig. \ref{figure-feyn} (c) and Fig. \ref{figure-feyn} (d). These $2\rightarrow 2$ scatterings (and various other channels) can be directly  tested at present \cite{coupling-LHC} and future  \cite{coupling-FCC,coupling-ILC} colliders.

\subsubsection{Right-Handed Neutrinos}
Symmergence has testable implications also for right-handed neutrinos. Indeed, if the bound (\ref{sm-NP-coup}) is to be respected and if the active neutrinos are to have correct masses ($m_{\nu}\lesssim 1\ {\rm eV}$)  then the right-handed neutrinos must have masses \cite{neutrino-portal,vissani}
\begin{eqnarray}
m_{N^{\prime}} \lesssim 1000\, {\rm TeV}
\label{mass-bound}
\end{eqnarray}
which can be probed at future experiments (based presumably on Higgs factories and accelerator neutrinos \cite{biz})  if not indirectly at the near-future SHiP experiment \cite{SHiP}. For $m_{\nu}\lesssim 0.1\ {\rm eV}$ \cite{sunny}, as revealed with recent cosmological data,  the bound in (\ref{mass-bound}) increases by an order of magnitude.

\subsection{Dark matter weighs below weak scale}
Symmergence has candidates for missing matter in both the ebony and dark NP sectors. The question of which one is preferred by nature can be answered only by observations (on, for instance, less-baryonic galaxies like cosmic seagull \cite{seagull}). 

The dark NP can be probed directly \cite{cdm-current,dm-LHC}. The scalar field $H^{\prime}$ in (\ref{action-sm-NP-int}), for instance, develops a  vanishing VEV, $\langle H^{\prime}\rangle = 0$, and becomes stable if it is real and has odd ${\mathbb{Z}}_2$ parity with respect to $H$. It is stable but its density is depleted by not only the expansion of the Universe but also the  coannihilations into the SM fields as $H^{\prime}H^{\prime}\rightarrow W^+ W^-, Z Z, \bar{t} t, \cdots$. The coannihilations into $W/Z$ are Goldstone-equivalent to $H^{\prime}H^{\prime}\rightarrow H H$ in Fig. \ref{figure-feyn} (b) so that $H^{\prime}$ scalars attain the observed relic density \cite{cdm-current,scalar-dm} if
\begin{eqnarray}
\lambda_{H H^{\prime}}^4 \left(\frac{{\rm GeV}}{m_{H^{\prime}}}\right)^2 \approx 10^{-7}
\end{eqnarray}
or, impliedly,
\begin{eqnarray}
m_{H^{\prime}} \lesssim 367\ {\rm GeV}
\label{dark-bound}
\end{eqnarray}
after using the seesawic structure in (\ref{sm-NP-coup}) for $\lambda_{H H^{\prime}}$. This bound on $m_{H^{\prime}}$, which makes sense barely as it lies beneath $\Lambda_W$, implies that thermal dark matter cannot weigh above the electroweak scale. (Non-thermal dark matter, belonging presumably to the ebony NP, is free from this bound.)  This means that symmergence can be probed by measuring the dark matter mass \cite{cdm-current,dm-LHC} and contrasting it with (\ref{dark-bound}) or, more safely, with $\Lambda_W$.

\section{Conclusion}
\begin{figure}[ht]
\includegraphics[scale=0.54]{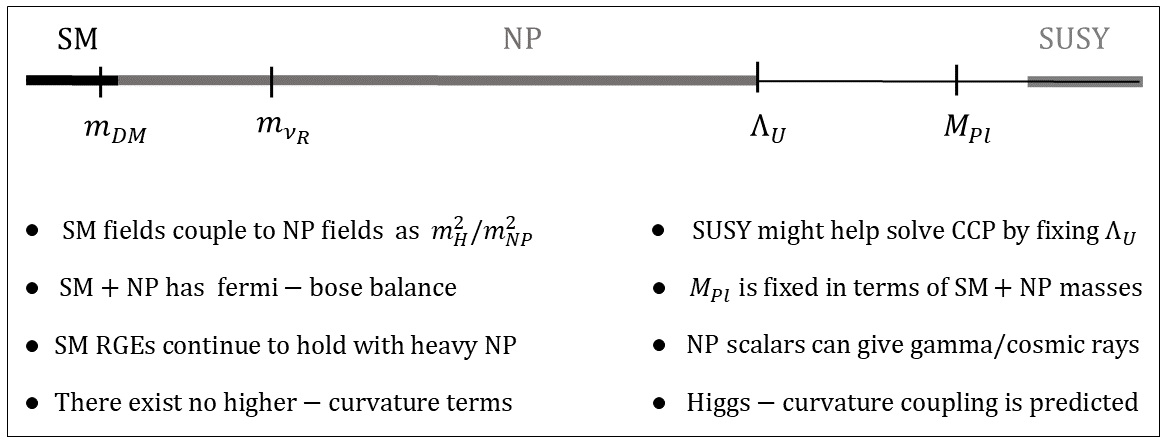}
\caption{\label{figure-summ} Fundamental aspects of the symmergent GR plus seesawic NP setup.}
\end{figure}

Needless to repeat,  symmergence propounds a novel framework in which notorious problems of the SM can be consistently addressed. It incorporates GR into the SM with a seesawic NP sector, and can be probed conclusively via various experimental tests ranging from collider searches to dark matter. Highlighted in Fig. \ref{figure-summ} are the salient features of the symmergent GR plus seesawic NP setup. It is clear that symmergence with a seeswic NP leads to a natural and viable setup with various testable predictions. 

The mechanism can be furthered in various aspects. First,  the trans-Planckian SUSY, its breaking mechanism, and its possible role in solving the CCP need be explored properly. Second, the MAG with quadratic curvature terms (\ref{action-affine-2px}) can give  GR-like geometry in small curvature limit and conformal affine geometry in high curvature limit so that a complete solution \cite{talk-qdis} can shed light on strong gravity regime (like black holes). Third, $Q_{\mu\nu}$ in the MAG action  (\ref{action-affine-2px}) is suggestive of bimetric gravity \cite{bimetric}, whose investigation may provide alternative views  of symmergence. Fourth, $Q_{\mu\nu}$ in the MAG action  (\ref{action-affine-2px}) is reminiscent of the disformal curvature couplings \cite{disformal}, whose exploration may shed light on the cosmological implications of symmergence. These four and various other aspects need be studied in detail to have a complete picture of symmergence. Experimental tests are of paramount importance to decide if symmergence is realized in nature or not. 

This work is supported in part by T{\"U}B{\.I}TAK grants 115F212 and 118F387. The author is grateful to Hemza Azri, Dieter van den Bleeken and Tekin Dereli for fruitful discussions on especially the emergent MAG. He thanks S. Vagnozzi and H. Terazawa for useful e-mail and mail exchanges.  

%\section*{References}

\end{document}